\documentclass[review,12pt,sort&compress]{elsarticle}

\usepackage{lineno,hyperref}
\journal{Digital Signal Processing}
\usepackage{graphicx}
\usepackage{amsfonts}
\usepackage{mathrsfs}
\usepackage{amssymb,amsmath}
\usepackage{bm}
\usepackage{algorithm}
\usepackage{algorithmic}
\usepackage{theorem}
\usepackage{array,color}
\usepackage{stfloats}
\usepackage{subfigure}
\usepackage{array}
\usepackage{multirow}
\usepackage{bbding}
\usepackage{longtable}
\begin{document}
\begin{frontmatter}

\title{TOA-based Passive Localization of Multiple Targets with Inaccurate Receivers Based on Belief Propagation on Factor Graph}
\author{Nan Wu, Weijie Yuan, Hua Wang$^*$ and Jingming Kuang}
\address{School of Information and Electronics, Beijing Institute of Technology, Beijing, China}
\cortext[mycorrespondingauthor]{Corresponding author}
\ead{\{wunan, wjyuan, wanghua, jmkuang\}@bit.edu.cn}
%
%
%
%
%
%
%
%
%
%
%
%
%
%
%
%

\begin{abstract}
Location awareness is now becoming a vital requirement for many practical applications. In  this  paper, we consider passive localization of multiple targets with one transmitter and several receivers based on time of arrival (TOA) measurements. Existing studies assume that positions of receivers are perfectly known. However, in practice, receivers' positions might be inaccurate, which leads to localization error of targets. We propose factor graph (FG)-based belief propagation (BP) algorithms to locate the passive targets and improve the position accuracy of receivers simultaneously. Due to the nonlinearity of the likelihood function, messages on the FG cannot be derived in closed form. We propose both sample-based and parametric methods to solve this problem. In the sample-based BP algorithm, particle swarm optimization is employed to reduce the number of particles required to represent messages. In parametric BP algorithm, the nonlinear terms in messages are linearized, which results in closed-form Gaussian message passing on FG. The Bayesian Cram\'{e}r-Rao bound (BCRB) for passive targets localization with inaccurate receivers is derived to evaluate the performance of the proposed algorithms. Simulation results show that both the sample-based and parametric BP algorithms outperform the conventional method and attain the proposed BCRB. Receivers' positions can also be improved via the proposed BP algorithms. Although the parametric BP algorithm performs slightly worse than the sample-based BP method, it could be more attractive in practical applications due to the significantly lower computational complexity.
\end{abstract}
\begin{keyword}
Passive Localization, Time of Arrival, Localization of Multiple Targets, Inaccurate Receivers' Positions, Factor Graph, Belief Propagation, Sample based method, Parametric Message Passing
\end{keyword}

\end{frontmatter}
\section{Introduction}
For many applications in wireless networks such as public service, emergence rescue, intelligent transportation system and environmental monitoring, it is often a crucial requirement to obtain the locations of targets \cite{gezici2005localization}. Depending on the nature of the object to be localized, localization methods can be classified into ``active'' and ``passive''. In passive localization, targets can only reflect or scatter the signals from transmitter to receivers \cite{friedlander1987passive}. Compared to the active case, there are increasing demands for localization of passive objects in many scenarios, such as crime-prevention, health care service and radar tracking.

Depending on how the distances are measured, localization techniques can be classified into three kinds, utilizing time of arrival (TOA), angle of arrival (AOA) and received signal strength (RSS) \cite{mao2007wireless,peng2006angle,lin2013accurate}. AOA methods rely on the equipment of directional antennas or antenna arrays at the receiver side. The accuracy of RSS methods suffers from the fading of wireless signal. Generally speaking, TOA algorithms can provide highly accurate estimation of target's position in most situations. Therefore, in this work, we focus on the TOA-based passive localization.

Various estimation methods based on TOA measurement have been presented for localization of active target.
Exact and approximate maximum likelihood (ML) localization algorithms are proposed in \cite{chan2006exact}. In \cite{guvenc2009survey}, numerous TOA-based wireless localization algorithms with different accuracies, computational complexities are surveyed. In \cite{wymeersch2009cooperative} and \cite{patwari2005locating}, cooperative localization algorithms are presented to provide high accuracy localization in anchorless network. A source localization method with channel estimation and noise reduction is studied in \cite{park2011block}. Using Bayesian framework, \cite{sathyan2013fast} presents a cooperative tracking method.
For passive localization in wireless networks, a TOA-based two-step estimation (TSE) algorithm is proposed in \cite{shen2012accurate}. Passive localization in quasi-synchronous networks is studied in \cite{6746730} based on TSE algorithm. In \cite{zhou2011indoor}, differential TOA measurements are employed to perform passive localization in asynchronous network. All the above methods assume that positions of transmitter and receivers are accurately known. However, in large-scale network or emergency-deployed situation, equipping all the receivers with location reference devices may be time cost and energy prohibitive. Therefore, receivers' position information may not be accurate in this case. Obviously, utilizing the erroneous position information of receivers directly would result in large positioning error. Many papers have considered the localization in the presence of transceiver's position uncertainty. In \cite{srirangarajan2008distributed}, a second-order cone programming (SOCP)-based algorithm is proposed. An expectation maximization-based approach is presented in \cite{bin2013em} which provides closed-form expressions of nodes' location coordinates. In \cite{taylor2006simultaneous}, a Bayesian filter is utilized to provide on-line probabilistic estimates. A distributed ML algorithm is proposed in \cite{kantas2012distributed} to jointly locate the sensor and target nodes. In \cite{li2015localization}, a source localization problem is solved in the presence of sensor location errors. In \cite{meyer2012simultaneous} and \cite{savic2013simultaneous}, the graph model based algorithms are proposed to solve the simultaneous self-localization and tracking in cooperative localization scenarios. However, all of these papers focus on active localizations. In our prior work \cite{yuan2014passive}, passive localization of a single target with inaccurate receivers is studied based on Gaussian message passing.

In this paper, we study the TOA-based passive localization of multiple targets in the presence of inaccurate receivers, which is an extension of the single target problem. We propose two algorithms based on belief propagation (BP) on factor graph (FG) \cite{kschischang2001factor}. Due to the nonlinearity of the likelihood function, the messages cannot be derived in closed form. To this end, we propose both sample-based and parametric methods to represent messages on FG. Particle swarm optimization (PSO) \cite{kennedy2010particle} is employed in the sample-based BP algorithm to reduce the number of particles required. In the parametric BP algorithm, the nonlinear term in messages are linearized by Taylor expansion. Accordingly, all the messages can be represented in Gaussian closed form, which results in Gaussian message passing \cite{loeliger2007factor} on FG. We derive Bayesian Cram\'{e}r-Rao bound (BCRB) for location estimation in this problem. Finally, the performance of the proposed algorithms are evaluated by Monte Carlo simulations.

The remainder of this paper is organized as follows. The problem formulation and system model are given in Section II. In Section III, factor graph for passive localization of multiple targets in the presence of inaccurate receivers is given. Both the PSO enhanced sample-based BP and the parametric BP with Gaussian message passing are proposed. The BCRB for position estimations of targets and receivers is derived in Section IV. Simulation results and discussions are given in Section V. Finally, conclusions are drawn in Section VI.

\emph{Notation}: $(\cdot)^T$, $(\cdot)^{-1}$ are the transpose and inverse operator, respectively; $\|\cdot\|$ is the Euclidean norm; $\mathbb {E}$ denotes the expectation operator; $\nabla_x$ is the differential operator with respect to $x$; $\textrm{diag}\{\bm{x}\}$ represents a diagonal matrix with main diagonal entries being the elements of $\bm{x}$ and the entries outside the main diagonal are all zero; $\bm{A}\succeq \bm{B}$ denotes that $\bm{A}-\bm{B}$ is positive semi-definite. We also denote by $\mu_{f\to x}$ the message from factor node to variable node and $\mu_{x\to f}$ the message from variable node to factor node.

\section{System Model and Problem Formulation}
\begin{figure}[!t]
\centering
\includegraphics[width=.8\textwidth]{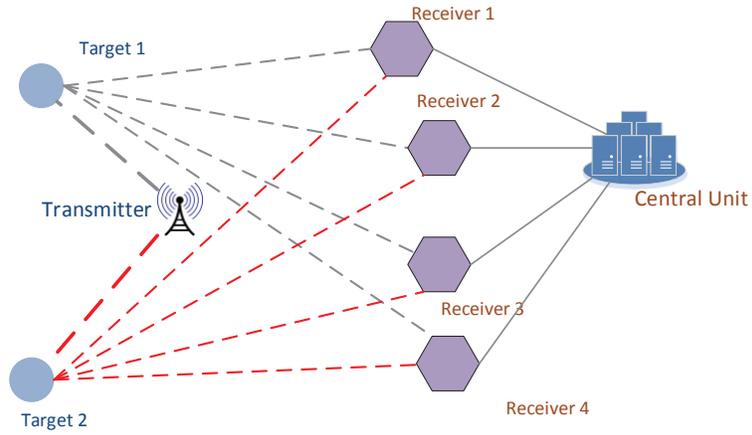}
\caption{\quad A multiple targets passive localization network}\label{Fig1}
\centering
\end{figure}

We consider a two-dimensional localization problem as shown in Fig.~\ref{Fig1}. The network consists of one transmitter, $A=|{\cal {A}}|$ passive targets and $M=|{\cal {M}}|$ receivers, where $\cal{A}$ and $\cal{M}$ denote the set of targets and receivers, respectively. Without loss of generality, the location of transmitter is set to $(0,0)$. The position of the $i$-th target and the $m$-th receiver is $\bm{x}_i=(x_i,y_i)$ and $\bm{\theta}_m=(a_{m},b_m)$, respectively. It is assumed that the transmitter and receivers are synchronized.

The transmitter sent an impulse at time $t$, which is then reflected by the $i$-th target and received by the ${m}$th receiver at time $t+t_{im}$, where $t_{im}$ denotes the signal propagation time from transmitter to the $m$-th receiver through the reflection of the $i$-th target. Signals reflected by different targets can be separated according to the blind source separation method and data association \cite{19535529,blackman2004multiple}. In this paper, we assume that signals have been perfectly separated for simplicity. Multiplying $t_{im}$ by the signal propagation speed $c$, the range measurement from transmitter via the $i$-th target to the $m$-th receiver can be written as
\begin{align}\label{range}
{R}_{im}=c\cdot t_{im} +n_{im}=\sqrt{x_i^2+y_i^2}+\sqrt{(x_i-{a}_m)^2+(y_i-{b}_m)^2}+n_{im},
\end{align}
where $n_{im}$ is a zero-mean Gaussian random variable with variance $\sigma_{im}^2$. We can write the likelihood function as
\begin{align}\label{LLF}
&p(R_{im}|\bm{x}_i,\bm{\theta}_m)\\
=&\frac{1}{\sqrt{2\pi\sigma_{im}^2}}\exp\left(\frac{\left(R_{im}-\sqrt{x_i^2+y_i^2}-\sqrt{(x_i-{a}_m)^2+(y_i-{b}_m)^2}\right)^2}{2\sigma_{im}^2}\right).\nonumber
\end{align}

For simplicity, let the vector ${\mathbf{x}}=[\bm{x}_1^T,..,\bm{x}_A^T]^T$ denote the positions of targets and ${\mathbf{\varTheta}}=[\bm{\theta}_1^T,...,\bm{\theta}_M^T]^T$ denote the receivers' positions. The vector ${\bm{R}}=[R_{11},...,R_{im},...,R_{AM}]^T, i\in{\cal A},m\in {\cal M}$ denotes all the range measurements. The joint likelihood function is given by
\begin{align}\label{likelihood}
p\left({\bm{R}}|{\mathbf{x}},\mathbf{\varTheta}\right)=\prod_{i\in{\cal A}, m\in{\cal M}} p(R_{im}|\bm{x}_i,\bm{\theta}_m).
\end{align}

We aim to obtain positions of targets based on the minimum mean square error (MMSE) estimator by considering the position uncertainties of receivers. Using Bayesian rule, the joint posterior distribution $p\left({\mathbf{x}},{\varTheta}|{\bm{R}}\right)$ reads
\begin{align}\label{joint_posteriori}
    p\left({\mathbf{x}},{\varTheta}|{\bm{R}}\right)&\propto p\left({\bm{R}}|{\mathbf{x}},{\varTheta}\right)p({\mathbf{x}})p\left({\varTheta}\right)
\\
    &=\prod_{i,m}p\left(R_{im}|\bm{x}_i,\bm{\theta}_m\right)\prod_{i\in{\cal A}}p\left(\bm{x}_i\right)\prod_{m\in{\cal M}}p(\bm{\theta}_m),\nonumber
\end{align}
where $p\left(\bm{x}_i\right)$ and $p(\bm{\theta}_m)$ are the prior distributions of targets and receivers, respectively. Then, the posterior distributions of the $i$-th target and the $m$-th receiver can be obtained by marginalizing the joint distribution in \eqref{joint_posteriori}, i.e.,
\begin{align}\label{marginal_function}
    p\left(\bm{x}_i|{\bm{R}}\right)&\propto\iint  p\left({\mathbf{x}},{\varTheta}|{\bm{R}}\right) \textrm{d}{\varTheta} \,\textrm{d}{\mathbf{x}}_{\sim\bm{x}_i} ,\\ \label{marginal_function_1}
    p\left(\bm{\theta}_m|{\bm{R}}\right)&\propto\iint  p\left({\mathbf{x}},{\varTheta}|{\bm{R}}\right) \textrm{d}{\mathbf{x}} \,\textrm{d}{\varTheta}_{\sim\bm{\theta}_m},
\end{align}
where '${\mathbf{x}}_{\sim\bm{x}_i}$' and '${\varTheta}_{\sim\bm{\theta}_m}$' denote that the integration is over all variables except the one in the subscript.

Due to the high dimensional integration, calculating the marginal distributions in \eqref{marginal_function} and \eqref{marginal_function_1} directly are difficult. In the following, we propose to use BP algorithm on FG to solve the problem efficiently.

%


\section{The Proposed Belief Propagation Algorithms on Factor Graph}

\subsection{Factor Graph and Message Passing Algorithm}

\begin{figure}[!t]
\centering
\includegraphics[width=.75\textwidth]{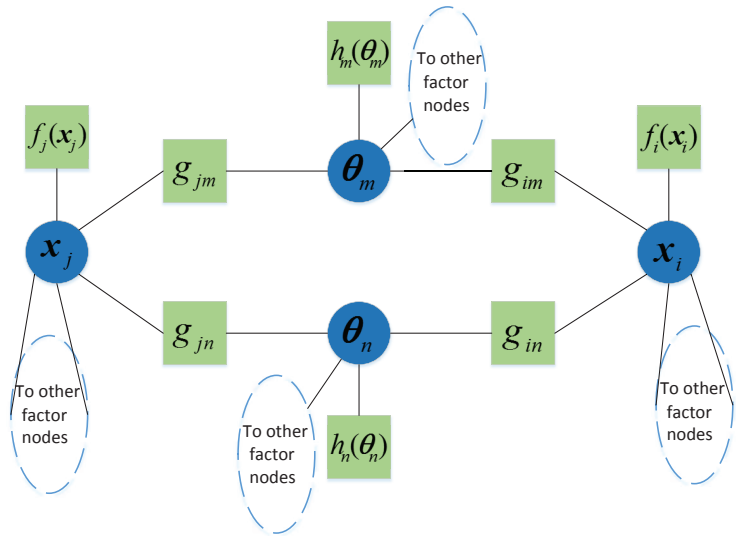}
\caption{\quad Factor graph of the passive localization system. Factor nodes $f$ and $h$ denote the prior distributions and node $g$ denotes the likelihood function.}\label{Fig2}
\centering
\end{figure}

Factor graph is a way to graphically show the factorization of a function \cite{kschischang2001factor}. In a factor graph, there is a factor node for every local function and a variable node for each variable. The factor node $g$ is connected with a variable node $x$ if and only if $g$ is a function of $x$. The factor graph corresponding to the factorization in \eqref{joint_posteriori} is shown in Fig. \ref{Fig2}, where $f_i$ and $h_m$ denote the prior distributions of $\bm{x}_i$ and $\bm{\theta}_m$, $g_{im}$ denotes the likelihood function $p(R_{im}|\bm{x}_i,\bm{\theta}_m)$. Based on the factor graph, the marginal posterior distribution $p(\bm{x}_i|\bm{R})$ and $p(\bm{\theta}_m|\bm{R})$ can be approximated by the beliefs $b(\bm{x}_i)$ and $b(\bm{\theta}_m)$ using belief propagation rules.

In BP message passing, there are messages from variable nodes to factor nodes and messages from factors to variable nodes \cite{pearl1986fusion}. The messages from factor node $g_{im}$ to variable nodes $\bm{x}_i$ and $\bm{\theta}_m$ are given by
\begin{align}\label{gtox}
\mu_{g_{im}\to \bm{x}_i}(\bm{x}_i)=&\int p(R_{im}|\bm{x}_i, \bm{\theta}_m)\mu_{\bm{\theta}_m \to g_{im}}(\bm{\theta}_m) d\bm{\theta}_m,\\\label{gtotheta}
\mu_{g_{im}\to \bm{\theta}_m}(\bm{\theta}_m)=&\int p(R_{im}|\bm{x}_i, \bm{\theta}_m)\mu_{ \bm{x}_i \to g_{im}}( \bm{x}_i) d \bm{x}_i,
\end{align}
where $\mu_{ \bm{x}_i \to g_{im}}( \bm{x}_i)$ and $\mu_{\bm{\theta}_m \to g_{im}}(\bm{\theta}_m)$ are messages from variable nodes to factor node $g_{im}$ which are given by
 \begin{align}\label{xtog}
 \mu_{ \bm{x}_i \to g_{im}}( \bm{x}_i)&=\prod_{s\in{\cal M}\backslash m}\mu_{g_{is}\to \bm{x}_i}(\bm{x}_i),\\\label{thetatog}
  \mu_{\bm{\theta}_m \to g_{im}}(\bm{\theta}_m)&=\prod_{k\in{\cal A}\backslash i}\mu_{g_{km}\to \bm{\theta}_m}(\bm{\theta}_m).
 \end{align}

The belief of a variable node is obtained by multiplying all the messages passed from neighboring factors. Hence, the belief $b(\bm{x}_i)$ and $b(\bm{\theta}_m)$ are given by
\begin{align}\label{bx}
b(\bm{x}_i)&=f_i(\bm{x}_i)\prod_{s\in{\cal M}}\mu_{g_{is}\to \bm{x}_i}(\bm{x}_i),\\ \label{btheta}
b(\bm{\theta}_m)&=h_m(\bm{\theta}_m)\prod_{k\in{\cal A}}\mu_{g_{km}\to \bm{\theta}_m}(\bm{\theta}_m).
\end{align}

The prior distribution of receivers is assumed to be circular symmetric Gaussian distribution, i.e.,
\begin{align}\label{circular}
p(\bm{\theta}_m)=\frac{1}{\sqrt{2\pi\sigma_{m}^2}}\exp\left(-\frac{\|\bm{\theta}_m-\bar{\bm{\theta}}_m\|^2}{2\sigma_{m}^2}\right),
\end{align}
where $\bar{\bm{\theta}}_m=(\bar{a}_m,\bar{b}_m)$ is the true position of the $m$-th receiver and $\sigma^2_{m}$ is the variance. Similarly, the prior distribution of targets $p\left(\bm{x}_i\right)$ can also be assumed as Gaussian\footnote{If there is no prior information of targets, $p\left(\bm{x}_i\right)$ is set to uniform distribution, which can be regarded as Gaussian with infinite variance.}. However, due to the nonlinearity of likelihood function, no close-form solutions can be obtained from the integration in \eqref{gtox} and \eqref{gtotheta}. We propose two methods, namely, sample-based and parametric methods, to solve this problem.

\subsection{Sample-based Belief Propagation Algorithm for Passive Localization}


The integration in \eqref{gtox} and \eqref{gtotheta} can be solved by Monte Carlo method, e.g., particle filtering \cite{ihler2005nonparametric}. For any integrable function $\phi(\bm{x})$, the integral $I=\int \phi(\bm{x}) r(\bm{x}) d\bm{x}$ can be approximated by using $L$ weighted particles, i.e., $\{\bm{x}^{(j)},\omega^{(j)}\}_{j=1}^L$, sampled from the distribution $r(\bm{x})$, i.e.,
$$I\approx\sum_{l=1}^L \omega^{(j)} \phi(\bm{x}^{(j)}),$$
where $\omega^{(j)}$ is the weight of the $j$-th particle.

Therefore the messages in \eqref{gtox} and \eqref{gtotheta} can be represented by the particles $\{\bm{x}_i^{(j)},\omega^{(j)}_i\}_{j=1}^L$ and $\{\bm{\theta}_m^{(k)},\omega^{(k)}_m\}_{k=1}^M$. The weights and values of particles can be obtained by using importance sampling \cite{978374}. After obtaining the particle representation of a message, the weights of particles are updated as
\begin{align}\label{updateweight}
\omega^{(j)}_i=\prod_{s\in{\cal M}}\mu_{g_{is}\to \bm{x}_i}(\bm{x}^{(j)}_i).
\end{align}
Then, the weights are normalized, i.e., $\omega^{(j)}_i=\omega^{(j)}_i/\sum_{j=1}^{L}\omega^{(j)}_i$. Resampling can be used to mitigate the degeneracy of particles. However, it also leads to the loss of diversity of particles. We propose to employ PSO method to solve this problem. PSO aims to find both the local best and the global best positions of a group of particles. The position and velocity of the $j$-th particle in a PSO iteration is updated as
\begin{align}\label{PSO}
\bm{x}_i^{(j)(q)}=&\bm{x}_i^{(j)(q-1)}+\bm{v}_i^{(j)(q)},\\\label{PSOposition}
\bm{v}_i^{(j)(q)}=&\bm{v}_i^{(j)(q-1)}+c_1 \left(\bm{p}_i^{(j)(q-1)}-\bm{x}_i^{(j)(q-1)}\right)+c_2  \left(\bm{g}_i^{(q-1)}-\bm{x}_i^{(j)(q-1)}\right),
\end{align}
where $c_1$ and $c_2$ are constants that control the convergence speed \cite{eberhart1995new}, $\bm{p}_i^{(j)(q-1)}$ is the local best position of the $j$-th particle and $\bm{g}_i^{(q-1)}$ is the global best position of all the particles, which are updated as follows
\begin{align}  \label{localbest}
\bm{p}_i^{(j)(q)}&=\left\{
\begin{aligned}
&{\bm{x}_i^{(j)(q)}} &   &\text{if} ~~J({\bm{x}_i^{(j)(q)}})\geq J({\bm{p}_i^{(j)(q-1)}})\\
&{\bm{p}_i^{(j)(q-1)}} &  &\text{otherwise} \\
\end{aligned}
\right.
\\\label{globalbest}
\bm{g}_i^{(q)}&=\arg\max_{j}\left(J({\bm{p}_i^{(j)(q)}})\right),
\end{align}
where $J(\cdot)$ is the objective function, which is set to be the likelihood function in this paper. The particle representation and PSO process of receivers' positions $\bm{\theta}_m$ follow the similar rule and are not given here for brevity. With the particle representation of the beliefs, the positions of targets and receivers can be obtained. The proposed PSO enhanced sample-based BP passive localization algorithm is depicted in \textbf{Algorithm} \ref{algorithm1}.

\subsection{Parametric Belief Propagation Algorithm for Passive Localization}

The estimation accuracy of sample-based BP algorithm depends on the number of particles. However, a large number of particles leads to high computational complexity. To this end, we propose a parametric BP algorithm for passive localization which significantly reduces the computational complexity.

\begin{figure}[!t]
\centering
\includegraphics[width=.7\textwidth]{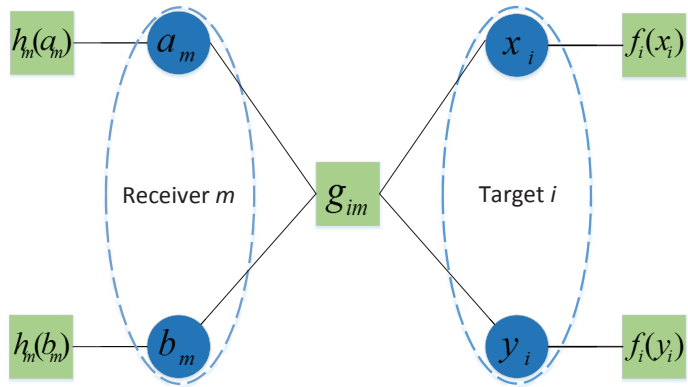}
\caption{\quad Subgraph of pairwise nodes in Fig. \ref{Fig2}}\label{Fig3}
\centering
\end{figure}

Assuming that $x$-axis and $y$-axis are independent, the message from factor node to variable node can be separated into two messages. A subgraph of pairwise nodes $\bm{x}_i$ and $\bm{\theta}_m$ is illustrated in Fig. \ref{Fig3}. The messages from factor nodes $f_i$ and $h_m$ to variable nodes are the prior distributions of the variable nodes. The messages from factor node $g_{im}$ to variable nodes are
\begin{align}\label{mugtox}
\hspace{-5mm}\mu_{g_{im}\to x_i}(x_i)\!\propto\!\iiint& \!p(R_{im}|x_i,y_i,a_m,b_m)\nonumber\\\times&\mu_{a_m\to g_{im}}(a_m)
\mu_{b_m\to g_{im}}(b_m)\mu_{y_i\to g_{im}}(y_i)\,\mathrm{d} a_m \,\mathrm{d} b_m \,\mathrm{d} y_i,\\
\hspace{-6mm}\mu_{g_{im}\to y_i}(y_i)\!\propto\!\iiint& \!p(R_{im}|x_i,y_i,a_m,b_m)\nonumber\\\times&\mu_{a_m\to g_{im}}(a_m)
\mu_{b_m\to g_{im}}(b_m)\mu_{x_i\to g_{im}}(x_i)\,\mathrm{d} a_m \,\mathrm{d} b_m \,\mathrm{d} x_i,\\
\hspace{-5mm}\mu_{g_{im}\to a_m}(a_m)\!\propto\!\iiint& \!p(R_{im}|x_i,y_i,a_m,b_m)\nonumber\\\times&\mu_{x_i\to g_{im}}(x_i)
\mu_{y_i\to g_{im}}(y_i)\mu_{b_m\to g_{im}}(b_m)\,\mathrm{d} x_i \,\mathrm{d} y_i \,\mathrm{d} b_m,\\\label{mugtoyi}
\hspace{-5mm}\mu_{g_{im}\to b_m}(b_m)\!\propto\!\iiint& \!p(R_{im}|x_i,y_i,a_m,b_m)\nonumber\\\times&\mu_{x_i\to g_{im}}(x_i)
\mu_{y_i\to g_{im}}(y_i)\mu_{a_m\to g_{im}}(a_m)\,\mathrm{d} x_i \,\mathrm{d} y_i \,\mathrm{d} a_m.
\end{align}


Substituting \eqref{LLF} into \eqref{mugtox}-\eqref{mugtoyi}, at the $l$-th iteration, we expand the terms of Euclidean norm in \eqref{mugtox}-\eqref{mugtoyi} based on the first order Taylor series around the position estimations of target $i$ and receiver $m$ in the $(l-1)$-th iteration, i.e., $(\hat{x}_i^{(l-1)},\hat{y}_i^{(l-1)})$ and $(\hat{a}_m^{(l-1)},\hat{b}_m^{(l-1)})$. By denoting $A_1,A_2,B_1,B_2$ as the directional derivatives on $x$-axis and $y$-axis, $A^{(l-1)}_1\triangleq\frac{\hat{x}_i^{(l-1)}-\hat{a}_m^{(l-1)}}{\hat{d}_{im}^{(l-1)}}$, $B^{(l-1)}_1\triangleq\frac{\hat{x}_i^{(l-1)}}{\hat{d}_i^{(l-1)}}$, $A^{(l-1)}_2\triangleq\frac{\hat{y}_i^{(l-1)}-\hat{b}_m^{(l-1)}}{\hat{d}_{im}^{(l-1)}}$, $B^{(l-1)}_2\triangleq\frac{\hat{y}_i^{(l-1)}}{\hat{d}_i^{(l-1)}}$, we have the approximations as
\begin{align}
\label{sqrt_approximate_agent}
&\sqrt{x_i^2+y_i^2}\approx\hat{d}_i^{(l-1)}+B^{(l-1)}_1\left(x_i-\hat{x}_i^{(l-1)}\right)+B^{(l-1)}_2\left(y_i-\hat{y_i}^{(l-1)}\right),\\
 \label{sqrt_approximate}
 &\sqrt{\left(x_i-a_m\right)^2+\left(y_i-b_m\right)^2}\approx\hat{d}_{im}^{(l-1)}+A^{(l-1)}_1\left(x_i-\hat{x}_i^{(l-1)}\right)
    +A^{(l-1)}_2\left(y_i-\hat{y}_i^{(l-1)}\right)
    \nonumber\\&\hspace{3cm}+A^{(l-1)}_1\left(\hat{a}_m^{(l-1)}-a_m\right)+A^{(l-1)}_2\left(\hat{b}_m^{(l-1)}-b_m\right),\\
 \label{sqrt multiple approximate}
&\sqrt{(x_i^2+y_i^2)\,\left(\left(x_i-a_m\right)^2+\left(y_i-b_m\right)^2\right)}\approx\hat{d}_i^{(l-1)}\Big(\hat{d}_{im}^{(l-1)}-A^{(l-1)}_1(a_m-\hat{a}_m^{(l-1)})\nonumber\\&\hspace{1.5cm}-A^{(l-1)}_2 (b_m-\hat{b}_m^{(l-1)})\Big)+(B_1^{(l-1)}\hat{d}_{im}^{(l-1)}+A_1^{(l-1)}\hat{d}_i^{(l-1)})(x_i-\hat{x}_i^{(l-1)})\nonumber\\&\hspace{1.5cm}+(B_2^{(l-1)}\hat{d}_{im}^{(l-1)}+A_2^{(l-1)}\hat{d}_i^{(l-1)})(y_i-\hat{y}_i^{(l-1)}),
\end{align}
where $\hat{d}_i^{(l-1)}$ and $\hat{d}_{im}^{(l-1)}$ are the estimated Euclidean distances in the previous iteration, $$\hat{d}_i^{(l-1)}\triangleq\sqrt{\left(\hat{x}_i^{(l-1)}\right)^2+\left(\hat{y}_i^{(l-1)}\right)^2},$$ $$\hat{d}_{im}^{(l-1)}\triangleq\sqrt{\left(\hat{a}_m^{(l-1)}-\hat{x}_i^{(l-1)}\right)^2+\left(\hat{b}_m^{(l-1)}-\hat{y}_i^{(l-1)}\right)^2}.$$

Assume the messages from variable nodes to factor nodes are Gaussian distributions. Substituting \eqref{sqrt_approximate_agent}-\eqref{sqrt multiple approximate} into \eqref{mugtox}-\eqref{mugtoyi}, after same straightforward manipulations, we can derive the messages from factor nodes to variable nodes as Gaussian distribution, i.e.,$$\mu^{(l)}_{g_{im} \to \xi}(\xi)={\cal N}\big(\xi,m_{g_{im}\to \xi}^{(l)},(\sigma_{g_{im}\to \xi}^{(l)})^2\big), \,\xi \in \{x_i, y_i, a_m, b_m\},$$
with the mean and variance given in Appendix 1.

Having all the messages from neighboring factor nodes, the beliefs of ${x}_i$, $y_i$, $a_m$ and $b_m$ can be obtained by
\begin{align}\label{b_x_i}
    b\left({x_i}\right)&=f_i(x_i)\prod_{s\in {\cal M}}\mu_{g_{is}\to x_i}\left({x_i}\right),\\b\left({y_i}\right)&=f_i(y_i)\prod_{s\in {\cal M}}\mu_{g_{is}\to y_i}\left({y_i}\right),\\\label{b_a_m}
  b\left({a_m}\right)&=h_m(a_m)\prod_{k\in {\cal A}}\mu_{g_{km}\to a_m}\left({a_m}\right),
\\ \label{b_b_m}
b\left({b_m}\right)&=h_m(b_m)\prod_{k\in {\cal A}}\mu_{g_{km}\to b_m}\left({b_m}\right).
\end{align}


Since all the messages from factor nodes to variable nodes are Gaussian, the beliefs in \eqref{b_x_i}-\eqref{b_b_m} are also Gaussian. The means and variances of $b^{(l)}(x_i)$ and $b^{(l)}(a_m)$ are
\begin{align}\label{bx1111}
m_{x_i}^{(l)}=&\left(\frac{1}{\left({\sigma}^{(0)}_{x_i}\right)^2}+\sum_{s\in{\cal M}}\frac{1}{\left(\sigma^{(l)}_{g_{is}\to x_i}\right)^2}\right)^{-1}\left(\frac{{m}^{(0)}_{x_i}}{\left({\sigma}^{(0)}_{x_i}\right)^2}+\sum_{s\in{\cal M}}\frac{m^{(l)}_{g_{is}\to x_i}}{\left(\sigma^{(l)}_{g_{is}\to x_i}\right)^2}\right),\\
\left(\sigma_{x_i}^{(l)}\right)^2=&\left(\frac{1}{\left({\sigma}^{(0)}_{x_i}\right)^2}+\sum_{s\in{\cal M}}\frac{1}{\left(\sigma^{(l)}_{g_{is}\to x_i}\right)^2}\right)^{-1},\\ \label{m_am}
m_{a_m}^{(l)}=&\left(\frac{1}{\left({\sigma}^{(0)}_{a_m}\right)^2}+\sum_{k\in{\cal A}}\frac{1}{\left(\sigma^{(l)}_{g_{km}\to a_m}\right)^2}\right)^{-1}\left(\frac{{m}^{(0)}_{a_m}}{\left({\sigma}^{(0)}_{a_m}\right)^2}+\sum_{k\in{\cal A}}\frac{m^{(l)}_{g_{km}\to a_m}}{\left(\sigma^{(l)}_{g_{km}\to a_m}\right)^2}\right),\\\label{byi111}
\left(\sigma_{a_m}^{(l)}\right)^2=&\left(\frac{1}{\left({\sigma}^{(0)}_{a_m}\right)^2}+\sum_{k\in{\cal A}}\frac{1}{\left(\sigma^{(l)}_{g_{km}\to a_m}\right)^2}\right)^{-1},
\end{align}
where ${\left({\sigma}^{(0)}_{x_i}\right)^2}$ and ${\left({\sigma}^{(0)}_{a_m}\right)^2}$ are the variances of the prior Gaussian distributions of targets and receivers in $x$-axis. The expressions of $b^{(l)}(y_i)$ and $b^{(l)}(b_m)$ have similar forms and are not given here for simplicity.

The messages from a variable node $\xi$ to its neighboring factor node $g$ can be calculated by the belief of $\xi$ divided by the incoming message from the factor node $g$ to the variable node $\xi$, i.e., $\mu_{\xi\to g}(\xi)=b(\xi)/\mu_{g\to \xi}(\xi)$, $\forall \xi\in\{x_i,y_i,a_m,b_m\}$. Therefore, the messages from variable node $\xi$ to its neighboring factor node $g$ at the $l$-th iteration are given by
\begin{align}\label{muxtof}
\mu^{(l)}_{\xi\to g}(\xi)\propto{\cal N}\Bigg(\xi,&\frac{{ m}^{(l)}_{\xi}\left(\sigma^{(l)}_{g\to \xi}\right)^2-m^{(l)}_{g\to \xi}\left(\sigma^{(l)}_{\xi}\right)^2}{\left(\sigma^{(l)}_{g\to \xi}\right)^2-\left(\sigma^{(l)}_{\xi}\right)^2},\frac{\left(\sigma^{(l)}_{\xi}\right)^2\left(\sigma^{(l)}_{g\to \xi}\right)^2}{\left(\sigma^{(l)}_{g\to \xi}\right)^2-\left(\sigma^{(l)}_{\xi}\right)^2}\Bigg),
\end{align}
$\forall \xi\in\{x_i,y_i,a_m,b_m\}.$

Since the FG in Fig.~\ref{Fig2} contains cycles, the above message update repeats until the number of iterations reaches the maximum. The proposed parametric BP algorithm for passive localization with inaccurate receivers is described in \textbf{Algorithm} \ref{algorithm3}.

\section{Analysis of Bayesian Cram\'{e}r-Rao bound}


The Cram\'{e}r-Rao bound (CRB) for passive localization with accurate receivers has been studied in \cite{shen2012accurate}. In this paper, since the positions of receivers are with uncertainties, we propose BCRB for joint estimation of positions of both targets and receivers.

Let $\bm{\rho}=[\bm{x}_1^T,\bm{x}_2^T,...,\bm{x}_A^T,\bm{\theta}_1^T,...,\bm{\theta}_M^T]^T$ be the vector of parameters to be estimated. For any unbiased estimator, the covariance matrix of $\hat{\bm{\rho}}$ satisfies $\textrm{cov}(\hat{\bm{\rho}})\succeq\bm{F}_\textrm{passive}^{-1}$, where $\bm{F}_\textrm{passive}$ is the Fisher information matrix (FIM), which is given by
\begin{align}\label{fisher}
\bm{F}_\textrm{passive}&=-\mathbb{E}\left[\nabla_{\bm{\rho}}\left\{\nabla_{\bm{\rho}}\left(\ln p(\bm{\rho}|\bm{R})\right)\right\}\right]\nonumber\\
      &=-\mathbb{E}\left[\nabla_{\bm{\rho}}\left\{\nabla_{\bm{\rho}}\left(\ln p(\bm{R}|\bm{\rho})\right)\right\}\right]-\mathbb{E}\left[\nabla_{\bm{\rho}}\left\{\nabla_{\bm{\rho}}\left(\ln p(\bm{\rho})\right)\right\}\right]\nonumber\\
      &=\bm{F}_\textrm{observ}+\bm{F}_\textrm{prior},
\end{align}
where $\bm{F}_\textrm{observ}$ and $\bm{F}_\textrm{prior}$ correspond to the contribution of observation and prior information to the FIM, respectively.

Assuming that the measurement noise is Gaussian distributed, the observation vector $\bm{R}$ is Gaussian distributed with mean vector $\bm{\mu}$ and covariance matrix $\bm{\Sigma}_1$, i.e.,
$$
\bm{\mu}=[\|\bm{x}_1-\bm{\theta}_1\|,\|\bm{x}_1-\bm{\theta}_2\|,...,\|\bm{x}_A-\bm{\theta}_M\|]^T,
$$
$$
\bm{\Sigma}_1=\textrm{diag}\{{\sigma^2_{11},\sigma^2_{12},\sigma^2_{13}...,\sigma^2_{AM}}\}.
$$
The matrix $\bm{F}_\textrm{observ}$ can be written as $\bm{F}_{\textrm{observ}}=\bm{J}_1\bm{\Sigma}_1^{-1}\bm{J}_1^{T}$, where $\bm{J}_1=\frac{\partial\bm{\mu}}{\partial\bm{\rho}}$ is the Jacobian matrix.

The prior distribution of $\bm{\rho}$ is also assumed as Gaussian distribution with mean vector at its true positions and the inversion of covariance matrix as
$$\bm{\Sigma}^{-1}_2=diag\{\bm{0}_{1\times 2A},\frac{1}{\sigma^2_{a1}},\frac{1}{\sigma^2_{b1}},...,\frac{1}{\sigma^2_{aM}},\frac{1}{\sigma^2_{bM}}\},$$ the FIM $\bm{F}_\textrm{prior}=\bm{J}_2\bm{\Sigma}_2^{-1}\bm{J}_2^{T}$ with $\bm{J}_2=\bm{I}_{2(A+M)\times 2(A+M)}$.

Therefore, $\bm{F}_{\textrm{passive}}$ can be written as
\begin{align}\label{FIM}
\bm{F}_{\textrm{passive}}=\bm{J}_1\bm{\Sigma}^{-1}_1\bm{J}_1^{T}+\bm{J}_2\bm{\Sigma}^{-1}_2\bm{J}_2^{T},
\end{align}
which can be expressed as a block matrix
\begin{align}\label{blockmatrix}
\bm{F}_{\textrm{passive}}=\left[
\begin{array}{cc}
\bm{F}_{11}&\bm{F}_{12}\\
\bm{F}_{21}&\bm{F}_{22}
\end{array}
\right],
\end{align}
where the derivations of $\bm{F}_{11}$, $\bm{F}_{12}$ and $\bm{F}_{22}$ are given in Appendix 2.

The BCRB is related to the inversion of matrix $\bm{F}_{\textrm{passive}}$. However, as the number of targets and receivers becomes larger, the computational complexity to calculate $\bm{F}_{\textrm{passive}}^{-1}$ increases significantly. We employ the equivalent Fisher information matrix (EFIM) to reduce the dimension of the FIM while retaining all necessary information related to targets or receivers \cite{13329033}.

We denote the EFIM corresponding to targets as $\bm{F}_{\textrm{target}}$. The EFIM $\bm{F}_{\textrm{target}}$ is the Schur complement of matrix ${\bm{F}_{22}}$, which is
\begin{align}\label{Ftarget}
\bm{F}_{\textrm{target}}=\bm{F}_{11}-\bm{F}_{12}{\bm{F}_{22}}^{-1}\bm{F}_{21}.
\end{align}
Obviously, calculating the inversion of $\bm{F}_{\textrm{target}}$ is much easier than that of $\bm{F}_{\textrm{passive}}$. Then, the BCRB for estimation of targets' positions $\mathbf{x}=[\bm{x}_1^T,\bm{x}_2^T,...,\bm{x}_A^T]^T$ is
\begin{align}\label{crlb}
\textrm{cov}(\hat{\bm{x}})\succeq \bm{F}_{\textrm{target}}^{-1}.
\end{align}
Similarly, the EFIM corresponding to receivers $\bm{F}_{\textrm{receiver}}$ can also be derived using the Schur complement of matrix ${\bm{F}_{11}}$, the expression of which is not given here for brevity.

\section{Simulation Results and Discussions}
\subsection{Simulation Results}
Consider a $100m\times 100m$ plane with one transmitter, three targets and five receivers. The transmitter is located at $(0,0)$. The positions of receivers are $(10,40)$, $(50,70)$, $(90,90)$, $(70,50)$, $(40,10)$, respectively. We assume that the prior distribution of position estimation errors of receivers are circular symmetric Gaussian distribution with zero-mean and the variance $\left(\sigma_m^{(0)}\right)^2$. The positions of targets are initialized to be uniformly distributed on the plane, which can be regarded as Gaussian distributions with infinite variances. The maximum number of iterations for BP on FG is set to $N_{iter}=40$ and the number of iterations for PSO is set to $N_{PSO}=10$. For simplicity, the same variance of range measurement noise $\sigma_{im}^2=1m^2$ is assumed for all the communication links. Similarly, we assume that the variances of receivers' prior distributions are identical, i.e., $\sigma^2=\left(\sigma_m^{(0)}\right)^2=9 m^2$, unless otherwise specified.

\begin{figure}[!t]
\centering
\includegraphics[width=.75\textwidth]{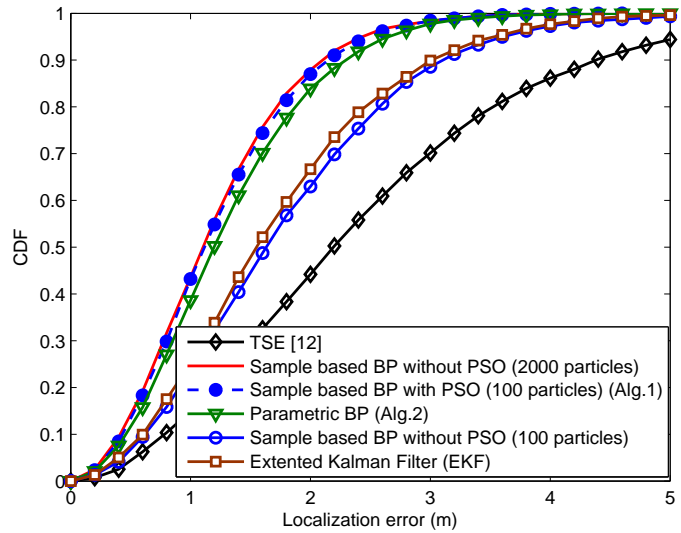}
\caption{\quad CDFs of targets localization error of different algorithms}\label{Fig4}
\centering
\end{figure}

\begin{figure}[!t]
\centering
\includegraphics[width=.75\textwidth]{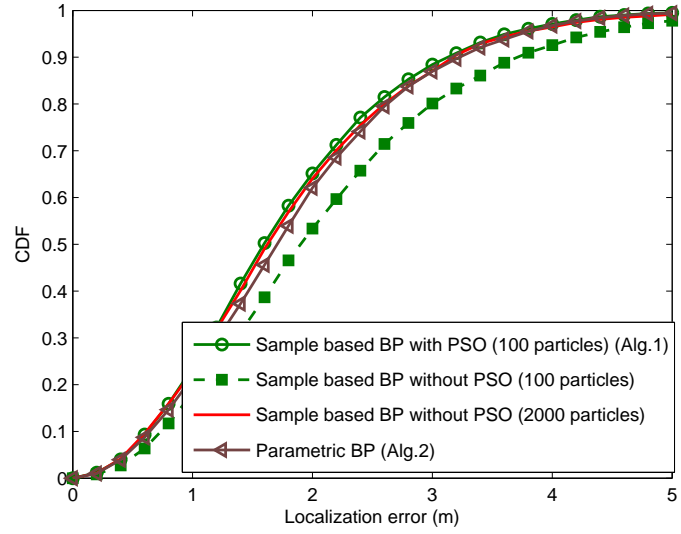}
\caption{\quad CDFs of receivers localization error of different algorithms}\label{cdfreceiver}
\centering
\end{figure}

The cumulative distribution functions (CDFs) of target localization error of the proposed algorithms are shown in Fig.~\ref{Fig4} and compared with that of the TSE method in \cite{shen2012accurate} and the extended Kalman filter (EKF). It is seen that the performance of TSE degrades significantly in the presence of inaccurate receivers. For EKF, since the the uncertainties of receivers are treated as additional measurement noise directly, performance loss can be observed. The proposed PSO enhanced sample-based BP algorithm with 100 particles performs very close to the one using 2000 particles without PSO. Therefore, by employing PSO, the number of particles can be notably reduced. This is because PSO can mitigate the loss of diversity of particles. Moreover, we can observe that the proposed PSO enhanced sample-based BP slightly outperforms the proposed parametric BP algorithm. However, since all the messages of the latter can be expressed in Gaussian close form, only the mean and variance have to be calculated, which results in much lower computational complexity. The CDFs of receivers' localization error using the proposed algorithms are illustrated in Fig.~\ref{cdfreceiver}. We can observe that the performance of receivers' position estimations based on the PSO enhanced sample-based BP and the parametric BP algorithms are close to each other.



\begin{figure}[!t]
\centering
\includegraphics[width=.75\textwidth]{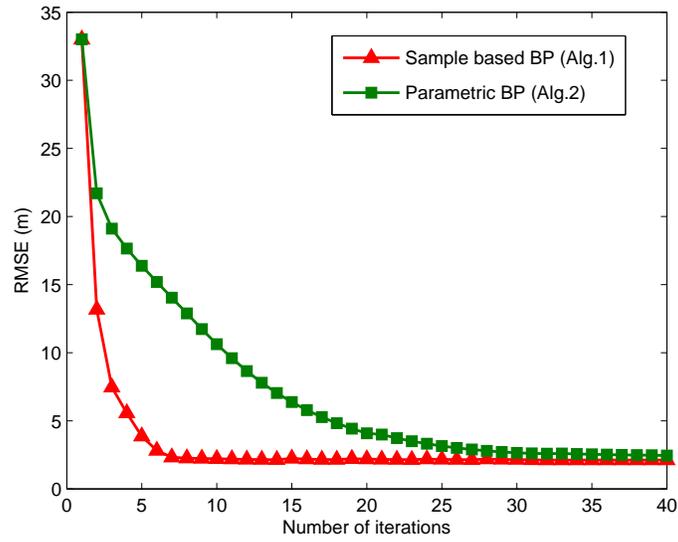}
\caption{\quad RMSE versus the number of iterations of the proposed algorithms}\label{Fig5}
\centering
\end{figure}

The root mean squared error (RMSE) of target's position estimation versus the number of iterations is illustrated in Fig.~\ref{Fig5}. We can observe that both the proposed sample-based BP and the parametric BP algorithm converge after several iterations. Due to the approximation in linearization, the parametric method converges slower than the sample-based BP algorithm. Nevertheless, the localization accuracies after convergence of the two algorithms are very close. Therefore, for space limitation, we will only evaluate the performance of parametric BP algorithm in the following.

\begin{figure}[!t]
\centering
\includegraphics[width=.75\textwidth]{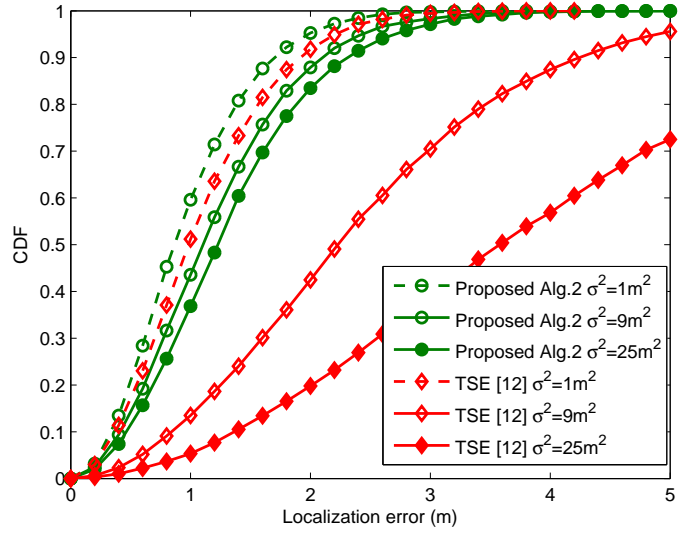}
\caption{\quad CDFs of targets localization error of different algorithms with various position uncertainties of receivers.}\label{Fig6}
\centering
\end{figure}

In Fig.~\ref{Fig6}, we compare the CDF of targets localization error of TSE method in \cite{shen2012accurate} with that of the proposed parametric BP algorithm with various position uncertainties of receivers. It is shown that, with larger position uncertainties of receivers, both the performance of TSE method and that of the proposed parametric BP algorithm degrade.  However, the performance gap between them becomes much larger as the uncertainty increases, which means by taking the position uncertainties of receivers into account, the proposed parametric BP algorithm for targets localization is more robust to the initial position errors of receivers.

\begin{figure}[!t]
\centering
\includegraphics[width=.75\textwidth]{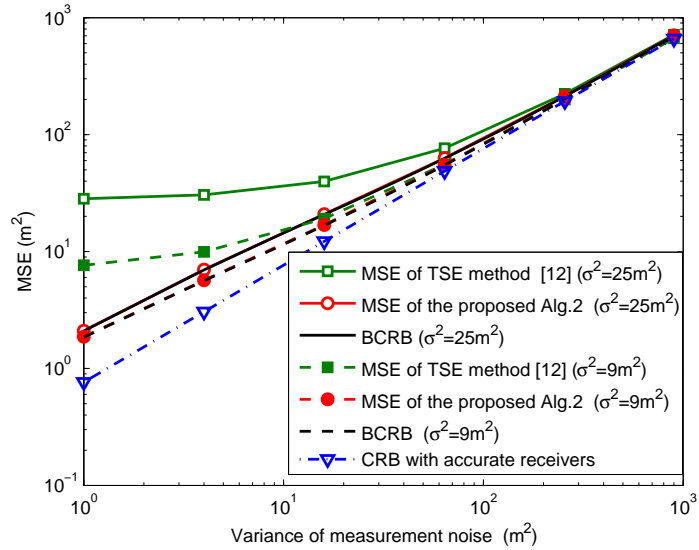}
\caption{\quad MSE of target localization versus the variance of measurement noise}\label{crlbfig}
\centering
\end{figure}

We evaluate the performance of the proposed algorithms by using the derived BCRB. Since it has been shown that the BCRB depends on the true positions of targets and receivers, we set the positions of the three targets at $(30,40),~(16,57)$ and $(70,81)$, respectively. Fig. \ref{crlbfig} illustrates the MSE of the position estimation of target located at $(30,40)$ using the proposed parametric BP algorithm and TSE method. The derived BCRBs with different receiver position uncertainties are also given for comparison. We can observe that the parametric BP algorithm can almost attain the derived BCRB, which verifies the efficiency of the proposed algorithm. It outperforms the TSE method at small measurement noise variance, where the receivers position uncertainties dominate the error of target localization.

\begin{figure}[!t]
\centering
\includegraphics[width=.75\textwidth]{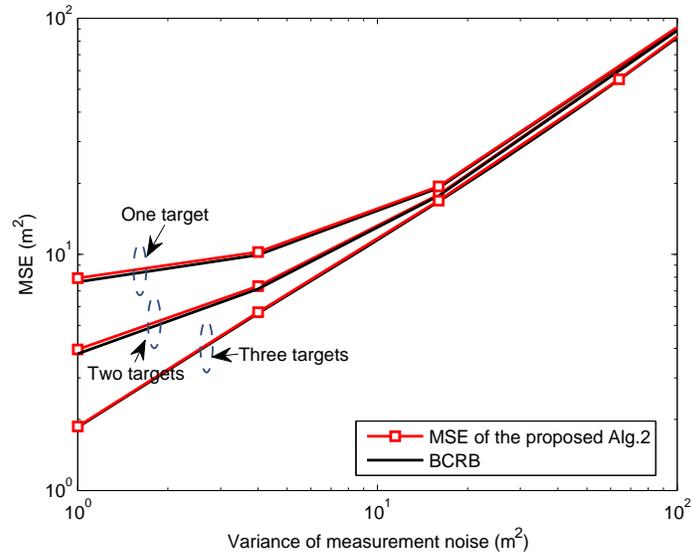}
\caption{\quad BCRB and MSE of the proposed target localization algorithm with different number of targets}\label{crlbnumber}
\centering
\end{figure}

The MSE of targets' positions estimation and the corresponding BCRB with different number of targets are evaluated in Fig.~\ref{crlbnumber}. Three configurations are considered, i.e., (1)one target located at $(30,40)$; (2)two targets located at $(30,40)$, $(16,57)$; (3)three targets located at $(30,40),~(16,57)$ and $(70,81)$. It is seen that localization accuracy of targets improves when the number of targets increases. In fact, from \eqref{b_a_m} and \eqref{b_b_m} we can observe that the belief of receiver's position combines the messages from the observations of multiple targets. Therefore, with greater number of targets, the accuracy of receiver's position can to be improved, which in turn leads to an improvement of the estimation accuracy of targets' positions. Moreover, it is seen that the proposed algorithm can attain the BCRB in all the three configurations.



\begin{figure}[!t]
\centering
\includegraphics[width=.75\textwidth]{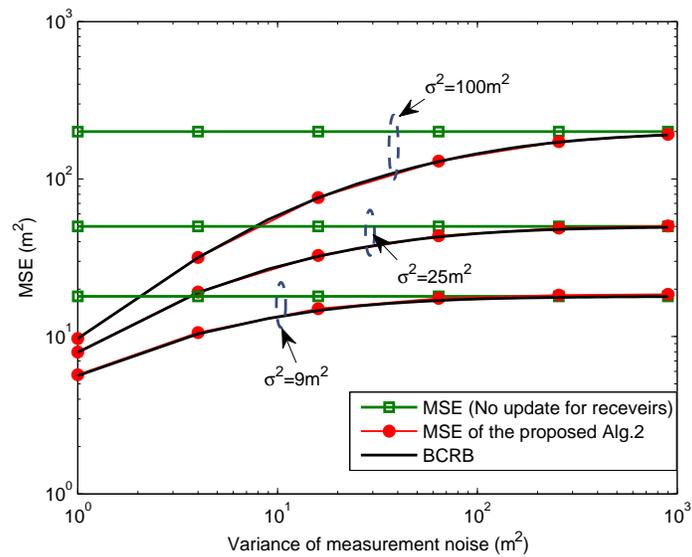}
\caption{\quad MSE of the receiver localization versus the variance of measurement noise}\label{crbreceiver}
\centering
\end{figure}

The BCRB and MSE of position estimation of receiver located at $(10,40)$ are plotted in Fig.~\ref{crbreceiver}. It is seen that the proposed parametric BP algorithm performs very close to the BCRB, which verifies the efficiency of the algorithm. For comparison purposes, the MSEs of receiver's initialized position estimation are also plotted, which is based on the realization of the prior distribution. We can observe that MSE of the proposed algorithm and the BCRB keep increasing when the variance of measurement noise becomes larger, and they converge to the initialized MSE of receiver's position estimation. This is because when the measurement noise is very large, the FIM of position estimation is dominated by the prior information. Moreover, it is seen that the BCRB and MSE of the proposed algorithm are below the MSE of the initial position estimation of receivers, which means that receivers can always benefit from the position updating using the proposed algorithm.

\begin{figure}[!t]
\centering
\includegraphics[width=.75\textwidth]{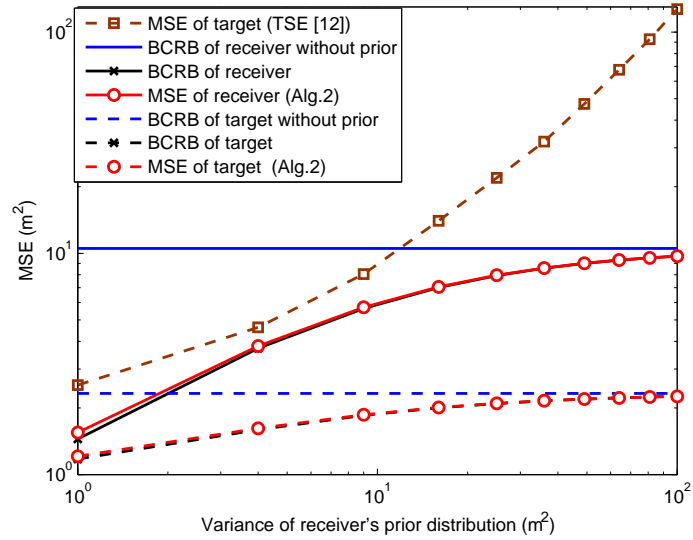}
\caption{\quad BCRB and MSE of the proposed algorithm versus the variance of receivers' prior distribution }\label{crbuncertain}
\centering
\end{figure}

To further evaluate the impact of receivers' position uncertainties, we plot BCRBs and MSEs of position estimations of both the target and the receiver versus the variance of prior distribution in Fig.~\ref{crbuncertain}. The variance of measurement noise is $\sigma^2_{im}=1m^2$. It is seen that, for target localization, when the variance of prior distribution is small, the MSE of TSE method converges to the BCRB which does not consider the prior information. When the variance of prior distribution increases, MSE of TSE method diverges quickly from that bound. In contrast, the proposed algorithm converges to that BCRB as the variance of prior distribution increases. It outperforms the TSE method and can almost attain the corresponding BCRB which takes the prior information into account. For receivers' location estimation, the MSE of the proposed algorithm performs close to its BCRB and converges to the bound which does not consider the prior information of receivers' position estimations.

%

\subsection{Computational Complexity Analysis}

We compare the computational complexity of the proposed sample-based BP and parametric BP algorithms. Assuming $R$ particles are used in the sample-based methods, the computational complexity of the sample-based BP algorithm without PSO scales as ${\cal O}((M+A)*{{R^2}})$, where $M$ and $A$ are the number of targets and receivers. PSO can help to reduce the number of particles in the sample-based BP at the cost of extra complexity of PSO iterations, and the total complexity scales as ${\cal O}((M+A)*{{R^2}})+{\cal O}(N_{PSO}*(M+A)*{{R}})$. Note that since PSO enhanced sample-based BP can reduce the number of particles, the value of $R$ in this case becomes much smaller, which results in lower complexity. For the proposed parametric BP algorithm, as only the means and variances of messages have to be calculated in each iteration, the complexity scales as ${\cal O}(M+A)$. The comparison of computational complexities of different algorithms are summarized in Table. \ref{complexity}.



\section{Conclusions}
In this paper, we proposed FG-based BP algorithms for passive localization of multiple targets in the presence of inaccurate receivers. To solve the intractable integrations in message updating on FG, a sample-based method was proposed to represent messages by particles. Moreover, PSO was employed to reduce the number of particles required, which resulted in lower computational complexity. In the parametric method, the nonlinear terms in the messages were linearized using Taylor expansions. Building on this approximation, all the messages on FG can be represented by Gaussian closed form, which lead to Gaussian message passing with means and variances to be updated. To evaluate the performance of the proposed algorithms, the BCRB of position estimations of both targets and receivers were derived. Simulation results showed that receivers' position uncertainties will lead to performance loss of targets localization. The proposed algorithms outperformed both the TSE and EKF method and can almost attain the BCRB. The proposed PSO enhanced sample-based BP algorithm can efficiently reduce the number of particles required. Due to the linearization approximation, the parametric BP algorithm performed slightly worse than the sample-based method. However, the computational complexity of the parametric BP was much lower than that of the sample-based method, which made the former more attractive in practical applications.

\section*{Acknowledgment}
This work is supported by ``National Science Foundation of China (NSFC)''(Grant Nos. 61201181, 61421001, 61471037) and ``A Foundation for the Author of National Excellent Doctoral Dissertation of P. R. China (FANEDD)''(Grant No. 201445).

\section*{Appendix}
\subsection*{\textbf{1.Means and variances of message $\mu_{g_{im}\to \xi}^{(l)}(\xi)$}}\label{appendix1}
The messages from variable nodes to factor nodes can be determined in Gaussian closed form as
$$\mu^{(l)}_{g_{im} \to \xi}(\xi)={\cal N}\big(\xi,m_{g_{im}\to \xi}^{(l)},(\sigma_{g_{im}\to \xi}^{(l)})^2\big),$$
with $\xi$ being the location coordinates of receivers and targets on $x$-axis and $y$-axis, the means and variances are given as follows,
\begin{align}\label{mugtox1}
m^{(l)}_{g_{im}\to x_i}=&\frac{1}{2\sigma_{im}^2+\left(\sigma_{{a_m\to g_{im} }}^{(l-1)}\right)^2} \Big[A_1\sigma_{im}^2{\left(R_{im}-\hat{d}_i^{(l-1)}\right)}\nonumber\\&\hspace{-2mm}+B_1\left(\sigma_{im}^2+\left(\sigma_{a_m\to g_{im} }^{(l-1)}\right)^2\right)\left(R_{im}-\hat{d}_{im}^{(l-1)}\right)+\sigma_{im}^2 m_{a_m\to g_{im}}^{(l-1)}\Big],\\
&\hspace{-5mm}(\sigma_{g_{im}\to x_i}^{(l)})^2=\frac{\sigma_{im}^2\left(\sigma_{im}^2+\left(\sigma_{a_m\to g_{im}}^{(l-1)}\right)^2\right)}{2\sigma_{im}^2+\left(\sigma_{a_m\to g_{im}}^{(l-1)}\right)^2},\\
m^{(l)}_{g_{im}\to y_i}=&\frac{1}{2\sigma_{im}^2+\left(\sigma_{{b_m\to g_{im} }}^{(l-1)}\right)^2}\Big[A_2\sigma_{im}^2{\left(R_{im}-\hat{d}_i^{(l-1)}\right)}\nonumber\\&\hspace{-2mm}+B_2\left(\sigma_{im}^2+\left(\sigma_{b_m\to g_{im} }^{(l-1)}\right)^2\right)\left(R_{im}-\hat{d}_{im}^{(l-1)}\right)+\sigma_{im}^2 m_{b_m\to g_{im}}^{(l-1)}\Big],\\
&\hspace{-5mm}(\sigma_{g_{im}\to y_i}^{(l)})^2=\frac{\sigma_{im}^2\left(\sigma_{im}^2+\left(\sigma_{b_m\to g_{im}}^{(l-1)}\right)^2\right)}{2\sigma_{im}^2+\left(\sigma_{b_m\to g_{im}}^{(l-1)}\right)^2,}\\
m^{(l)}_{g_{im}\to a_m}=&\frac{1}{\sigma_{im}^2+\left(\sigma_{x_i\to g_{im}}^{(l-1)}\right)^2} \Big[B_1\left(\sigma_{x_i\to g_{im}} ^{(l-1)}\right)^2{\left(R_{im}-\hat{d}_{im}^{(l-1)}\right)}\nonumber\\&\hspace{-2mm}-A_1\left(\sigma_{im}^2+\left(\sigma_{x_i\to g_{im}}^{(l-1)}\right)^2\right){\left(R_{im}-\hat{d}_i^{(l-1)}\right)}+\sigma_{im}^2 m_{x_i\to g_{im}}^{(l-1)}\Big],\\
&\hspace{-5mm}(\sigma_{g_{im}\to a_m}^{(l)})^2=\frac{\sigma^2_{im}\left(2\left(\sigma_{x_i\to g_{im}}^{(l-1)}\right)^2+\sigma^2_{im}\right)}{\left(\sigma_{x_i\to g_{im}}^{(l-1)}\right)^2+\sigma_{im}^2},\\
m^{(l)}_{g_{im}\to b_m}=&\frac{1}{\sigma_{im}^2+\left(\sigma_{y_i\to g_{im}}^{(l-1)}\right)^2} \Big[B_2\left(\sigma_{y_i\to g_{im}} ^{(l-1)}\right)^2{\left(R_{im}-\hat{d}_{im}^{(l-1)}\right)}\nonumber\\&\hspace{-2mm}-A_2\left(\sigma_{im}^2+\left(\sigma_{y_i\to g_{im}}^{(l-1)}\right)^2\right){\left(R_{im}-\hat{d}_i^{(l-1)}\right)}+\sigma_{im}^2 m_{y_i\to g_{im}}^{(l-1)}\Big],\\\label{mugtoyi1}
&\hspace{-5mm}(\sigma_{g_{im}\to b_m}^{(l)})^2=\frac{\sigma^2_{im}\left(2\left(\sigma_{y_i\to g_{im}}^{(l-1)}\right)^2+\sigma^2_{im}\right)}{\left(\sigma_{y_i\to g_{im}}^{(l-1)}\right)^2+\sigma_{im}^2}.
\end{align}
where $A_1$, $A_2$, $B_1$ and $B_2$ are directional derivatives, which have been defined previously.

\subsection*{\textbf{2. Derivation of Fisher information matrix $\bm{F}_{\textrm{passive}}$}}

$\bm{F}_{\textrm{passive}}$ can be represented in block matrix form as
\begin{align}\label{blockmatrix}
\bm{F}_{\textrm{passive}}=\left[
\begin{array}{cc}
\bm{F}_{11}&\bm{F}_{12}\\
\bm{F}_{21}&\bm{F}_{22}
\end{array}
\right],
\end{align}

$\bm{F}_{11}$ and $\bm{F}_{22}$ are block diagonal matrix with $A$ and $M$ sub-matrices, respectively. The $i$-th sub-matrix of $\bm{F}_{11}$ is
\begin{align}
{\bm{F}_{11}}_{i}=\left[
\begin{array}{cc}
\sum_{m=1}^M \frac{1}{\sigma_{im}^2} {A_{im}^2}&\sum_{m=1}^M \frac{1}{\sigma_{im}^2} {A_{im}\cdot B_{im}}
\\
\sum_{m=1}^M \frac{1}{\sigma_{im}^2} {A_{im}\cdot B_{im}}&\sum_{m=1}^M \frac{1}{\sigma_{im}^2} {B_{im}^2}
\end{array}
\right],
\end{align}
where the elements $A_{im}$, $B_{im}$, $C_{im}$ and $D_{im}$ are the partial derivatives given by
\begin{align} \nonumber
A_{im}\triangleq&\frac{x_i}{\sqrt{x_i^2+y_i^2}}+\frac{x_i-a_m}{\sqrt{(x_i-a_m)^2+(y_i-b_m)^2}}, \\ \nonumber B_{im}\triangleq&\frac{y_i}{\sqrt{x_i^2+y_i^2}}+\frac{y_i-b_m}{\sqrt{(x_i-a_m)^2+(y_i-b_m)^2}},\\ \nonumber
C_{im}\triangleq&\frac{a_m-x_i}{\sqrt{(x_i-a_m)^2+(y_i-b_m)^2}},\\ \nonumber
D_{im}\triangleq&\frac{b_m-y_i}{\sqrt{(x_i-a_m)^2+(y_i-b_m)^2}}.
\end{align}
The $m$-th sub-matrix of $\bm{F}_{22}$ is given by
\begin{align}
{\bm{F}_{22}}_{m}=\left[
\begin{array}{cc}
\sum_{i=1}^A \frac{1}{\sigma_{im}^2} {C_{im}^2}+\frac{1}{\sigma_{am}^2}& \sum_{i=1}^A \frac{1}{\sigma_{im}^2} {C_{im} \cdot D_{im}}
\\
\sum_{i=1}^A \frac{1}{\sigma_{im}^2} {C_{im}\cdot D_{im}}&\sum_{i=1}^A \frac{1}{\sigma_{im}^2} {D_{im}^2}+\frac{1}{\sigma_{bm}^2}
\end{array}
\right].
\end{align}
$\bm{F}_{12}=\bm{F}_{21}^T$ is the cross information between targets and receivers. $\bm{F}_{12}$ is a $A\times M$ block matrix, the sub-matrix in $i$-th row and $m$-th column is
\begin{align}
{\bm{F}_{12}}_{im}=\frac{1}{\sigma_{im}^2}\left[
\begin{array}{cc}
A_{im}\cdot C_{im}&A_{im}\cdot D_{im}
\\
B_{im}\cdot C_{im}&B_{im}\cdot D_{im}
\end{array}
\right].
\end{align}

\bibliography{IEEEabrv,bib}

\begin{thebibliography}{10}
\expandafter\ifx\csname url\endcsname\relax
  \def\url#1{\texttt{#1}}\fi
\expandafter\ifx\csname urlprefix\endcsname\relax\def\urlprefix{URL }\fi
\expandafter\ifx\csname href\endcsname\relax
  \def\href#1#2{#2} \def\path#1{#1}\fi

\bibitem{gezici2005localization}
S.~Gezici, Z.~Tian, G.~B. Giannakis, H.~Kobayashi, A.~F. Molisch, H.~V. Poor,
  Z.~Sahinoglu, Localization via ultra-wideband radios: a look at positioning
  aspects for future sensor networks, {IEEE} Signal Process. Mag. 22~(4) (2005)
  70--84.

\bibitem{friedlander1987passive}
B.~Friedlander, A passive localization algorithm and its accuracy analysis,
  {IEEE} J. Ocean. Eng. 12~(1) (1987) 234--245.

\bibitem{mao2007wireless}
G.~Mao, B.~Fidan, B.~Anderson, Wireless sensor network localization techniques,
  Comp. Netw. 51~(10) (2007) 2529--2553.

\bibitem{peng2006angle}
Q.~H. Spencer, B.~D. Jeffs, M.~A. Jensen, A.~L. Swindlehurst, Modeling the
  statistical time and angle of arrival characteristics of an indoor multipath
  channel, {IEEE} J. Sel. Areas Commun. 18~(3) (2000) 347--360.

\bibitem{lin2013accurate}
L.~Lin, H.-C. So, Y.~Chan, Accurate and simple source localization using
  differential received signal strength, Digit. Signal Process. 23~(3) (2013)
  736--743.

\bibitem{chan2006exact}
Y.-T. Chan, H.~Yau Chin~Hang, P.-c. Ching, Exact and approximate maximum
  likelihood localization algorithms, {IEEE} Trans. Veh. Technol. 55~(1) (2006)
  10--16.

\bibitem{guvenc2009survey}
I.~Guvenc, C.-C. Chong, A survey on TOA based wireless localization and NLOS
  mitigation techniques, {IEEE} Commun. Surveys Tuts. 11~(3) (2009) 107--124.

\bibitem{wymeersch2009cooperative}
H.~Wymeersch, J.~Lien, M.~Z. Win, Cooperative localization in wireless
  networks, Proc. {IEEE} 97~(2) (2009) 427--450.

\bibitem{patwari2005locating}
N.~Patwari, J.~N. Ash, S.~Kyperountas, A.~O. Hero, R.~L. Moses, N.~S. Correal,
  Locating the nodes: cooperative localization in wireless sensor networks,
  {IEEE} Signal Process. Mag. 22~(4) (2005) 54--69.

\bibitem{park2011block}
C.-H. Park, K.-S. Hong, Block LMS-based source localization using range
  measurement, Digit. Signal Process. 21~(2) (2011) 367--374.

\bibitem{sathyan2013fast}
T.~Sathyan, M.~Hedley, Fast and accurate cooperative tracking in wireless
  networks, {IEEE} Trans. Mobile Comput. 12~(9) (2013) 1801--1813.

\bibitem{shen2012accurate}
J.~Shen, A.~F. Molisch, J.~Salmi, Accurate passive location estimation using
  TOA measurements, {IEEE} Trans. Wireless Commun. 11~(6) (2012) 2182--2192.

\bibitem{6746730}
Y.~Wang, S.~Ma, C.~Chen, Toa-based passive localization in quasi-synchronous
  networks, {IEEE} Commun. Lett. 18~(4) (2014) 592--595.

\bibitem{zhou2011indoor}
Y.~Zhou, C.~L. Law, Y.~L. Guan, F.~Chin, Indoor elliptical localization based
  on asynchronous UWB range measurement, {IEEE} Trans. Instrum. Meas. 60~(1)
  (2011) 248--257.

\bibitem{srirangarajan2008distributed}
S.~Srirangarajan, A.~H. Tewfik, Z.-Q. Luo, Distributed sensor network
  localization using SOCP relaxation, {IEEE} Trans. Wireless Commun. 7~(12)
  (2008) 4886--4895.

\bibitem{bin2013em}
B.~Li, N.~Wu, H.~Wang, J.~Kuang, Expectation-maximization-based localization
  using anchors with uncertainties in wireless sensor networks, IET Commun.
  8~(11) (2014) 1977--1987.

\bibitem{taylor2006simultaneous}
C.~Taylor, A.~Rahimi, J.~Bachrach, H.~Shrobe, A.~Grue, Simultaneous
  localization, calibration, and tracking in an ad hoc sensor network, in:
  Proc. 5th Int. Conf. Inform. Process. Sensor Netw., 2006, pp. 27--33.

\bibitem{kantas2012distributed}
N.~Kantas, S.~S. Singh, A.~Doucet, Distributed maximum likelihood for
  simultaneous self-localization and tracking in sensor networks, {IEEE} Trans.
  Signal Process. 60~(10) (2012) 5038--5047.

\bibitem{li2015localization}
J.~Li, H.~Pang, F.~Guo, L.~Yang, W.~Jiang, Localization of multiple disjoint
  sources with prior knowledge on source locations in the presence of sensor
  location errors, Digit. Signal Process. 40 (2015) 181--197.

\bibitem{meyer2012simultaneous}
F.~Meyer, E.~Riegler, O.~Hlinka, F.~Hlawatsch, Simultaneous distributed sensor
  self-localization and target tracking using belief propagation and likelihood
  consensus, in: Proc. Forty Sixth Asilomar Conf. Signal. Sys. Comput.,
  2012, pp. 1212--1216.

\bibitem{savic2013simultaneous}
V.~Savic, H.~Wymeersch, Simultaneous localization and tracking via real-time
  nonparametric belief propagation, in: 2013 IEEE Int. Conf. Acoust., Speech
  and Signal Process., 2013, pp. 5180--5184.

\bibitem{yuan2014passive}
W.~Yuan, G.~Hao, N.~Wu, H.~Wang, J.~Kuang, Passive localization with inaccurate
  receivers based on gaussian belief propagation on factor graph, in: IEEE/CIC
  Conf. Commun. China, 2014, pp. 453--457.

\bibitem{kschischang2001factor}
F.~R. Kschischang, B.~J. Frey, H.-A. Loeliger, Factor graphs and the
  sum-product algorithm, {IEEE} Trans. Inf. Theory 47~(2) (2001) 498--519.

\bibitem{kennedy2010particle}
J.~Kennedy, Particle swarm optimization, in: Encyclopedia of Machine Learning,
  Springer, 2010, pp. 760--766.

\bibitem{loeliger2007factor}
H.-A. Loeliger, J.~Dauwels, J.~Hu, S.~Korl, L.~Ping, F.~R. Kschischang, The
  factor graph approach to model-based signal processing, Proc. {IEEE} 95~(6)
  (2007) 1295--1322.

\bibitem{19535529}
M.-Z. Poh, D.~J. McDuff, R.~W. Picard, {Non-contact, automated cardiac pulse
  measurements using video imaging and blind source separation}, Opt. Exp.
  18~(10) (2010) 10762--10774.

\bibitem{blackman2004multiple}
S.~S. Blackman, Multiple hypothesis tracking for multiple target tracking,
  {IEEE} Aerosp. Electron. Syst. Mag. 19~(1) (2004) 5--18.

\bibitem{pearl1986fusion}
J.~Pearl, Fusion, propagation, and structuring in belief networks, Artif.
  Intell. 29~(3) (1986) 241--288.

\bibitem{ihler2005nonparametric}
A.~T. Ihler, J.~W. Fisher, R.~L. Moses, A.~S. Willsky, Nonparametric belief
  propagation for self-localization of sensor networks, {IEEE} J. Sel. Areas
  Commun. 23~(4) (2005) 809--819.

\bibitem{978374}
M.~Arulampalam, S.~Maskell, N.~Gordon, T.~Clapp, A tutorial on particle filters
  for online nonlinear/non-Gaussian Bayesian tracking, {IEEE} Trans. Signal
  Process. 50~(2) (2002) 174--188.

\bibitem{eberhart1995new}
R.~C. Eberhart, J.~Kennedy, A new optimizer using particle swarm theory, in:
  Proc. the sixth Int. Symp. Micro Mach. Human Sci., 1995, pp. 39--43.

\bibitem{13329033}
Y.~Shen, M.~Z. Win, {Fundamental limits of wideband localization - part I: a
  general framework}, {IEEE} Trans. Inf. Theory 56~(10) (2010) 4956--4980.

\end{thebibliography}

\newpage

\begin{algorithm}[!t]
    \caption{Sample-based BP passive localization algorithm}
    \label{algorithm1}
    \begin{algorithmic}[1]
           \STATE {\bf{Initialization}}:~~{\bf{in parallel}}
        \STATE ~~~~target $i\in{\cal M}$ is initialized as circular symmetric Gaussian distribution \\~~~~with infinite variance;
        \STATE ~~~~{receiver} $m\in{\cal A}$ is initialized as circular symmetric Gaussian distribution;\\
        \STATE {\bf{end parallel}};
        \FOR{$l=1$ to $N_{iter}$}
              \STATE draw samples $\{\bm{x}_i^{(j)},\omega^{(j)}_i\}_{j=1}^L$ and $\{\bm{\theta}_m^{(k)},\omega^{(k)}_m\}_{k=1}^P$ from the beliefs $b^{(l-1)}(\bm{x}_i)$ and $b^{(l-1)}(\bm{\theta}_m)$.
              \STATE {\bf Particle swarm optimization (PSO) processing}
              \FOR{$q=1$ to $N_{PSO}$}
              \STATE update the positions and velocities of particles according to \eqref{PSO} and \eqref{PSOposition};
              \STATE update the local best position of each particle and the global best position according to \eqref{localbest} and \eqref{globalbest};
              \ENDFOR;
              \STATE {\bf end PSO}
              \STATE set ${\bm{x}_i^{(j)}}={\bm{p}_i^{(j)(N_{PSO})}}$
                  \STATE obtain the particle representation of messages \eqref{gtox} and \eqref{gtotheta} using samples $\{\bm{x}_i^{(j)},{\omega}^{(j)}_i\}_{j=1}^L$ and $\{\bm{\theta}_m^{(k)},{\omega}^{(k)}_m\}_{k=1}^M$.
                  \STATE update the weights of particles according to \eqref{updateweight} and do the normalization. Then obtain the particle representation of beliefs $b^{(l-1)}(\bm{x}_i)$ and $b^{(l-1)}(\bm{\theta}_m)$;
                  \STATE calculate the outgoing messages according to \eqref{xtog} and \eqref{thetatog};
                  \STATE estimate positions of targets and receivers based on MMSE criterion;
        \ENDFOR;
    \end{algorithmic}
\end{algorithm}

\begin{algorithm}[!t]
    \caption{Parametric BP passive localization algorithm}
    \label{algorithm3}
    \begin{algorithmic}[1]
    \STATE {\bf{Initialization}}: {\bf{in parallel}}
        \STATE target $i \in{\cal A}$ is initialized as Gaussian distribution with infinite variance;
        \STATE {receiver} $m\in{\cal M}$ is initialized as Gaussian distribution:
            \STATE ~~~$f_{a_m}\left(a_m\right)\propto{\cal N}\Big(a_m,m^{(0)}_{a_m},\big(\sigma^{(0)}_{a_m}\big)^2\Big)$,\\ ~~~$f_{b_m}\left(b_m\right)\propto{\cal N}\Big(b_m,m^{(0)}_{b_m},\big(\sigma^{(0)}_{b_m}\big)^2\Big)$, \\
        \STATE {\bf{end parallel}};
        \FOR{$l=1$ to $N_{iter}$}
              \STATE compute all messages from factor nodes to variable nodes according to \eqref{mugtox1}-\eqref{mugtoyi1}
                  \STATE update the means and variances of marginal beliefs $b^{(l)}\left(x\right)$, $b^{(l)}\left(y\right)$, $b^{(l)}\left(x_i\right)$, $b^{(l)}\left(y_i\right)$ according to \eqref{bx1111}-\eqref{byi111};
                 \STATE calculate the outgoing messages according to \eqref{muxtof} and send the messages to neighboring factor nodes;
                  \STATE estimate the positions of targets and receivers based on MMSE criterion;
        \ENDFOR;
    \end{algorithmic}
\end{algorithm}

\begin{table}[h]
\caption{Comparison of computational complexity of different algorithms}
\centering
\label{complexity}
\begin{tabular}{cc}
\hline
 Algorithm &  Computational Complexity \\
\hline
 Sample-based BP without PSO& ${\cal O}((M+A)*{{R^2}})$ \\
\hline
 Sample-based BP with PSO & ${\cal O}((M+A)*{{R^2}})+{\cal O}(N_{PSO}*(M+A)*{{R}})$  \\
\hline
Parametric BP & ${\cal O}(M+A)$  \\
\hline
\end{tabular}
\end{table}


\end{document}